\documentclass[11pt,twoside]{article}
\usepackage{asp2010}
\usepackage{natbib}
\usepackage{graphicx}
\usepackage{subfigure}
\usepackage{multirow}
\resetcounters

\begin{document}

\title{Photometric Observations of the Young Cluster Variable GM\,Cephei}
\author{S. C.-L. Hu$^{1,2}$, W. P. Chen$^{1,3}$, and the Young Exoplanet Transit Initiative
\affil{$^1$Institute of Astronomy, National Central University, 300 Jhongda Rd., Jhongli 32001, Taiwan}
\affil{$^2$Taipei Astronomical Museum, 363 Jihe Rd., Shilin, Taipei 11160, Taiwan}
\affil{$^3$Department of Physics, National Central University, 300 Jhongda Rd., Jhongli 32001, Taiwan}}

\begin{abstract}
We present our photometric observations of GM\,Cep, a solar type variable in the young ($\sim4$~Myr) open cluster
Trumpler~37.  The star is known as a classical T Tauri star with a circumstellar disk and active accretion.
GM\,Cep was suspected to undergo an outburst, thus a candidate for an EXor-type variable.  In our monitoring
campaign observations in 2010--2011, GM\,Cep experienced a $\sim0.82$~mag brightness decrease in the $R$ band
lasting for 39~days, and frequent, transient flare-like episodes with amplitude $\la 1$~mag, each lasting for
about 10~days.  The brightening was accompanied with a bluer color, presumably arising from increased
accretion activity.  Interestingly, the star also turned bluer in the fading phase.
Combining the AAVSO and literature data, we found a quasi-cyclic peroid of $\sim311$~days for the fading event.
A possible mechanism for the fading could be obscuration by a clump of dust around the star.  We proposed
that GM\,Cep therefore should be a UXor-type variable in the transition phase between grain coagulation and planetesimal
formation process in the circumstellar disk.
\end{abstract}

\section{Introduction}
\begin{figure}[h]
\begin{center}
\includegraphics[width=1\textwidth]{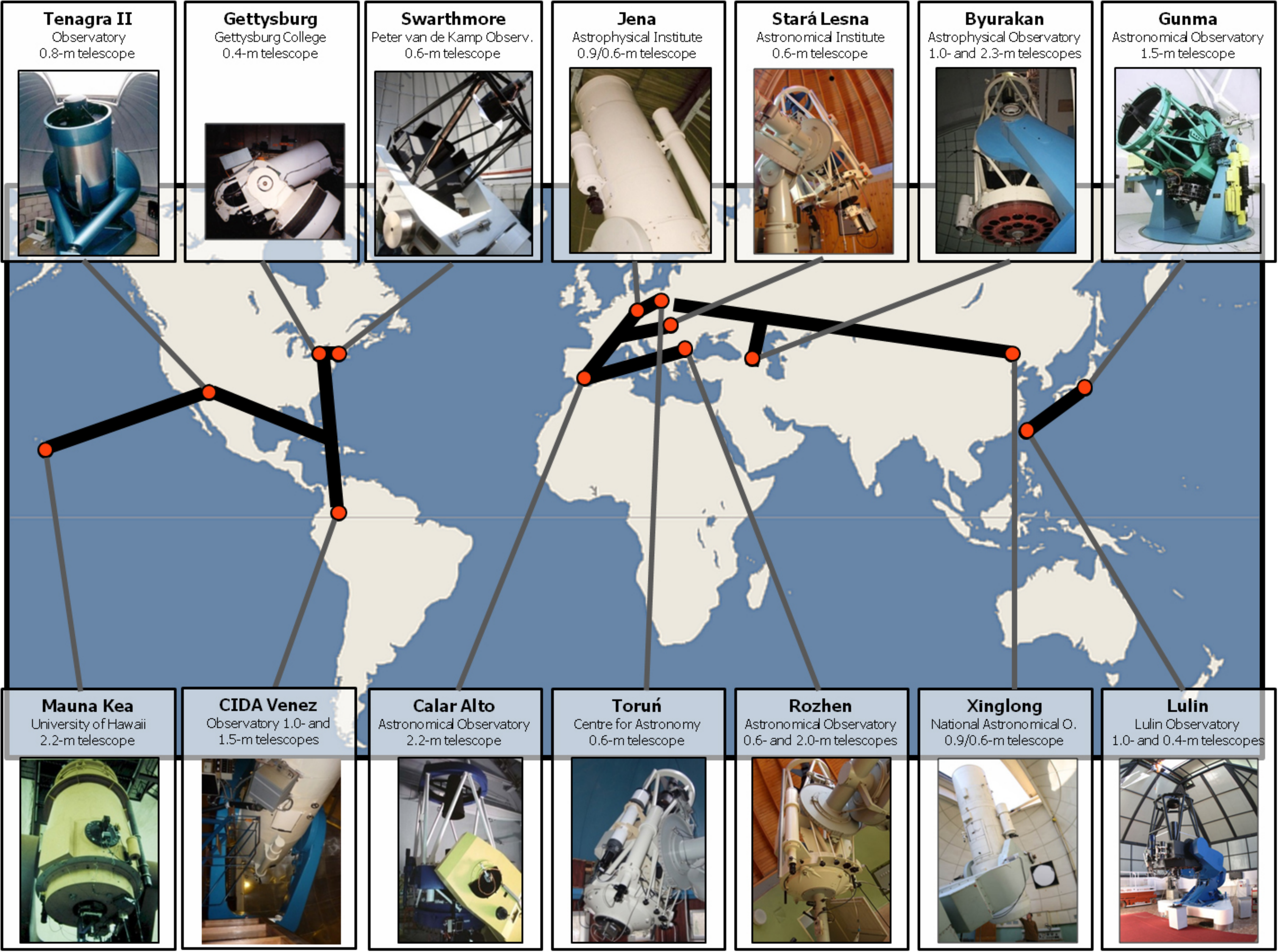}
\caption{The YETI network of telescopes. \label{fig1}}
\end{center}
\label{fig:1}
\end{figure}

The Young Exoplanet Transit Initiative (YETI) is an international multi-site project which consists of a
relay global network of telescopes (Fig.~\ref{fig:1}) to search for exoplanet transit events in young open clusters
\citep{neu11}.  During an observing campaign, telescopes at different longitude range are used to monitor a young, nearby
star cluster continuously for several days. Since 2009, the YETI has monitored two clusters, Trumpler (Tr) 37 at an age
of $\sim4$~Myr \citep{sic04} and 25\,Ori at an age of $\sim10$~Myr \citep{bri07}.  While the primary goal of the
YETI campaigns is to search for possible youngest exoplanets, the intensive monitoring produces data sets also
valuable for young stellar variability study, which might be relevant to planet formation \citep{bou03}.

The open cluster Tr\,37, at a heliodistance of 900~pc \citep{con02}, is part of the Cepheus
OB2 association.  Young stars are believed to evolve from Class II disk-bearing Classical T Tauri Stars (CTTSs)
to Class III diskless Weak-lined T Tauri Stars (WTTSs) in 1~Myr to 10~Myr, when
disk dissipation and planet formation are taking place.  With a disk frequency of $\sim39$\% \citep{mer09},
Tr\,37 hence serves as a good target to search for and to characerize exoplanets in formation or early evolutionary
stages \citep{neu11}.

Most pre-main sequences stars show irregular photometric variability.  \citet{her94} classified the variability into
three categories.  Type~I variation is a periodic modulation caused by the rotation of a star with cool spots.
Type~II variation is caused by unsteady accretion or the rotation of hot spots on the star surface.
Stars with Type~III variation, also called UXors, with UX\,Ori being the prototype, are hypothesized to
suffer variable obscuration by circumstellar dust.

GM\,Cep (RA = 21 38 17.3, Dec = +57 31 23, J2000) is a solar type variable in Tr\,37.  It has a possible spectral type
of G7 to K0, with an estimated mass of $2.1~M\sun$ and radius 3--6~$R\sun$ \citep{sic08}.  The youth of GM\,Cep is
exemplified by its emission-line spectra, prominent IR excess \citep{sic08}, and as an $Chandra$ X-ray source
\citep{mer09}, all characteristics of CTTSs.  It has a circumstellar disk \citep{mer09}, with an accretion rate
up to 10$^{-6}$~$M\sun$~yr$^{-1}$, which is 2--3 orders higher than the median value of the TTSs in Tr\,37
\citep{sic06}.  It is also one of the fastest rotators in the cluster, with $v \sin i \sim43.2$~km~s$^{-1}$ \citep{sic08}.

Early observations of GM\,Cep by the YETI in 2009 suggested a possible EXor-type outburst.
Previous studies on the light variability have been controversial.  \citet{sic08} collected photometric data of
GM\,Cep from 1952 to 2007 in the literature, supplemented by their own intensive multi-wavelength observations.
They suggested that GM\,Cep is an EXor type variable with an unstable disk and variable accretion rate.
On the other hand,
\citet{xia10} measured archival plates taken at Sonneberg and Harvard observatories between 1895 and 1993,
and concluded the variability in the long-term light curve to be dominated by dips (possibly from extinction)
superposed on quiescence states, instead of outbursts caused by accretion flares.  Here we present our multi-band
optical observations of GM\,Cep, and discuss the possible mechanism underlying its variability.

\section{Observations and Data Reduction}
Our optical observations in the $BVR_c$ bands were carried out with the 1.0~m and 0.4~m telescopes at the Lulin Observatory
(120.5E, 23.3N) in Taiwan and the 0.81~m telescope at the Tenagra Observatory (110.5W, 31.3N) in Arizona, USA.
The Lulin One-meter Telescope (LOT) was equipped with a Princeton Instruments model 1300B (PI-1300B) camera,
with a back-illuminated EEV CCD36-40 chip of $1340 \times 1300$~pixels.  Each 20 \micron\ pixel corresponds to
a plate scale of 0.51\arcsec per pixel, yielding a field of view of 11\arcmin$\times$11\arcmin.  For the YETI campaign,
 a 0.55$\times$ focal reducer was used to enlarge the field.  The 0.4~m SLT telescope at the Lulin Observatory was
equipped with an Apogee Alta U9000 front-illuminated CCD camera with a Kodak KAF-9000 sensor.  With
$3056 \times 3056$ pixels, each of 12~\micron\ on a side, the field of view is $37\arcmin \times 37\arcmin$.
Between 26 August to 22 November in 2010, another camera was used instead.  It was a back-illuminated Apogee Alta
U42 with E2V CCD42-40, which has $2048 \times 2048$ 13.5~\micron\ pixels, giving a field of view of $28\arcmin \times 28\arcmin$.
The 0.81~m Tenagra II telescope at the Tenagra Observatory was equipped with a back-illuminated SITe SI003 AP8p CCD,
with $ 1024 \times 1024$ 24~\micron\ pixels.  The field of view of the Tenagra II images is $ 15\arcmin \times 15\arcmin$.
All the photometry images were reduced following standard routines to correct the bias, dark and flat-field.
In addition, some data were taken with the 90/60~cm telescope of the University Observatory
Jena (11.5E, 50.9N). For imaging, the telescope works in the Schmidt mode, with an effective diameter of
0.6~m. It was equipped with E2V CCD42-10 (STK), with $2048 \times 2048$ 13.5~\micron\ pixels.  The images were
subtracted by overscan and dark, then divided by the master flat.

Aperture photometry was performed on GM\,Cep by comparison with the seven reference stars within 3\arcmin\ of
GM\,Cep used in \citet{xia10}.  These comparison stars have comparable brightness as that of GM\,Cep.  In each image we measured the FWHM of
GM\,Cep and the comparison stars, then 3$\times$, 5$\times$, and 7$\times$ the largest FWHM among them were used for
the aperture radius, inner sky radius, and outer sky radius, respectively.  A linear regression between the instrument magnitudes
and calibrated magnitudes of the comparison stars was utilized to compute the brightness of GM\,Cep.
Images with inferior sky conditions were excluded in the analysis.  In addition to our own
observations, visual observations from the American Association of Variable Star Observers (AAVSO) from 2006 onwards were collected
as well.

\section{Analysis}
\begin{figure}
\begin{center}
\mbox{
\subfigure[]{\includegraphics[width=0.475\textwidth]{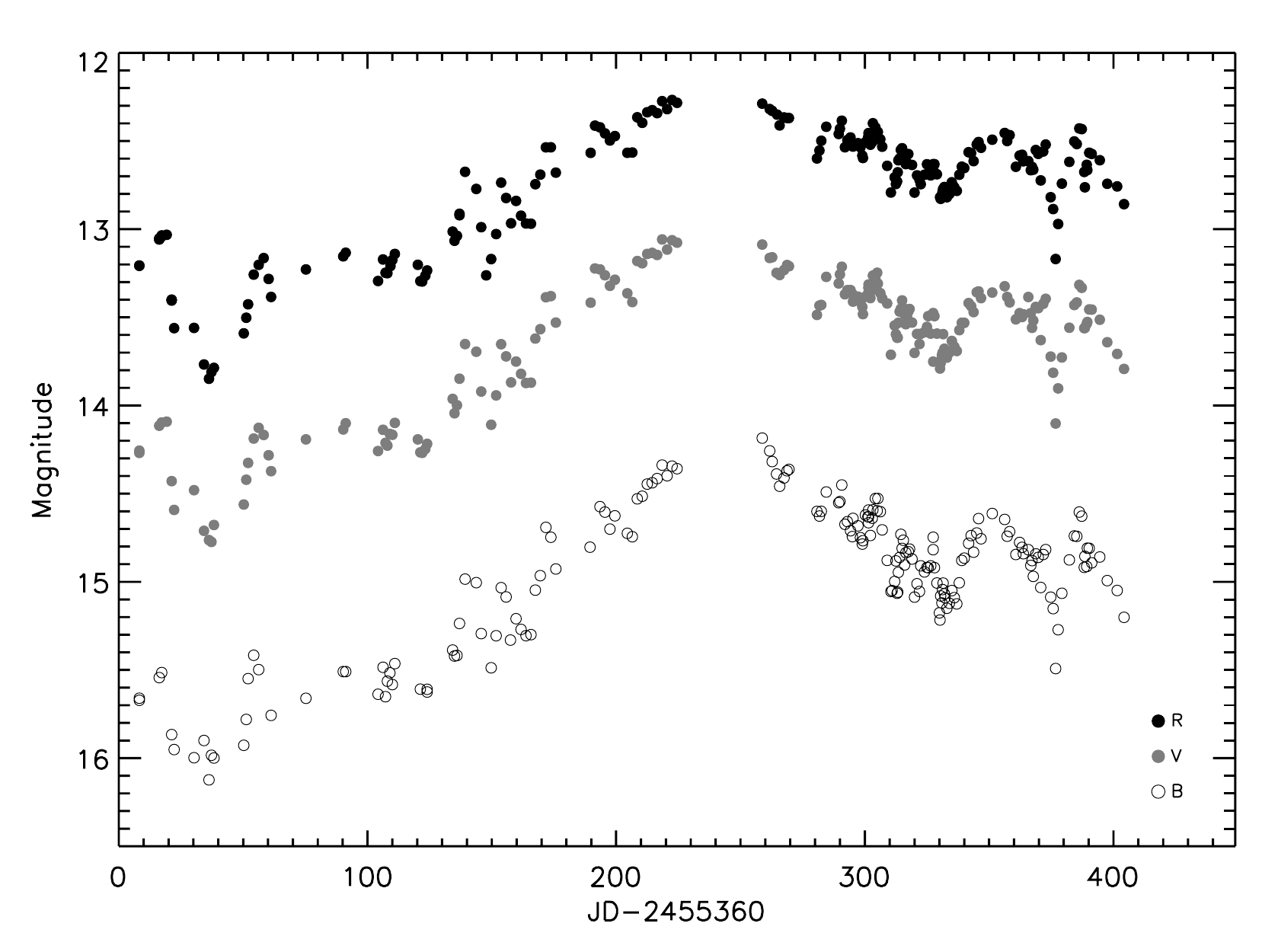}}}
\subfigure[]{\includegraphics[width=0.51\textwidth]{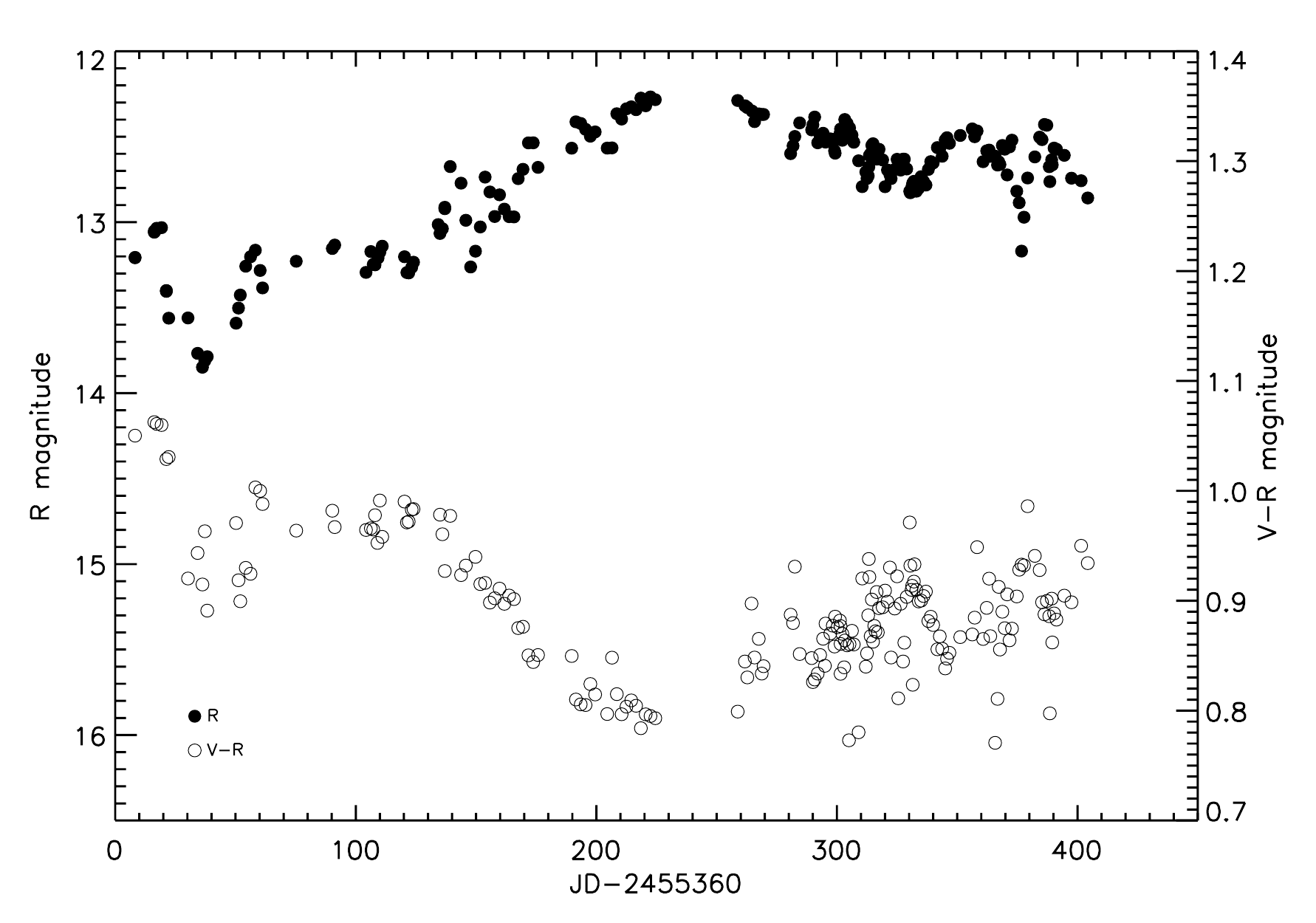}}
\caption{{\bf (a)} The light curves of GM\,Cep in the $B$ (open circles), $V$ (gray circles), and $R$ (filled circles) bands from our observations.  \label{fig2a}{\bf (b)} The $R$ band light curve (filled circles) and $V-R$ color curve (open circles) of GM\,Cep from our observations. \label{fig2b}}
\label{fig2}
\end{center}
\end{figure}

The light curves of GM\,Cep in the $B$, $V$, and $R$ bands are depicted in Fig.~\ref{fig2}(a).
Typical errors are smaller than the symbol sizes so are not shown.  The variations in different bands follow well with
each other, and abruptive fluctuations are obvious.  Several days after our monitoring campaign started in 2010,
GM\,Cep experienced a 0.82~mag fading in the $R$ band lasting for about 39~days.  The corresponding amplitude drop in the $V$ band
is smaller, about 0.68~mag.  One also sees transient flaring-like episodes with amplitude less than 1 mag, each lasting for about
10~days.  The nature of the variability is better diagnosed with color variations.  Our data in the $B$ band suffered large photometric
errors, so the $R$ band light curve and the $V-R$ color curve are presented in Fig.~\ref{fig2}(b).  One sees that the color
in general turned bluer as the star became brighter.  Interestingly, during the fading episode, GM\,Cep also had a bluer color.

We extended the time baseline of the light curve by including the data from the literature \citep{sic08} and from the AAVSO
archive.  A quasi-cyclic fading period of $\sim318$~days was discernible by eye inspection (Fig.~\ref{fig3}).
The NStED (NASA/IPAC/NExScI Star and Exoplanet Database) Periodogram Service was utilized to search for periodic signals with
the Lomb-Scargle algorithm, after removing the long-term trend in the light curve.
The first-ranked period is 311~days.

\begin{figure}
\begin{center}
\includegraphics[width=.84\textwidth]{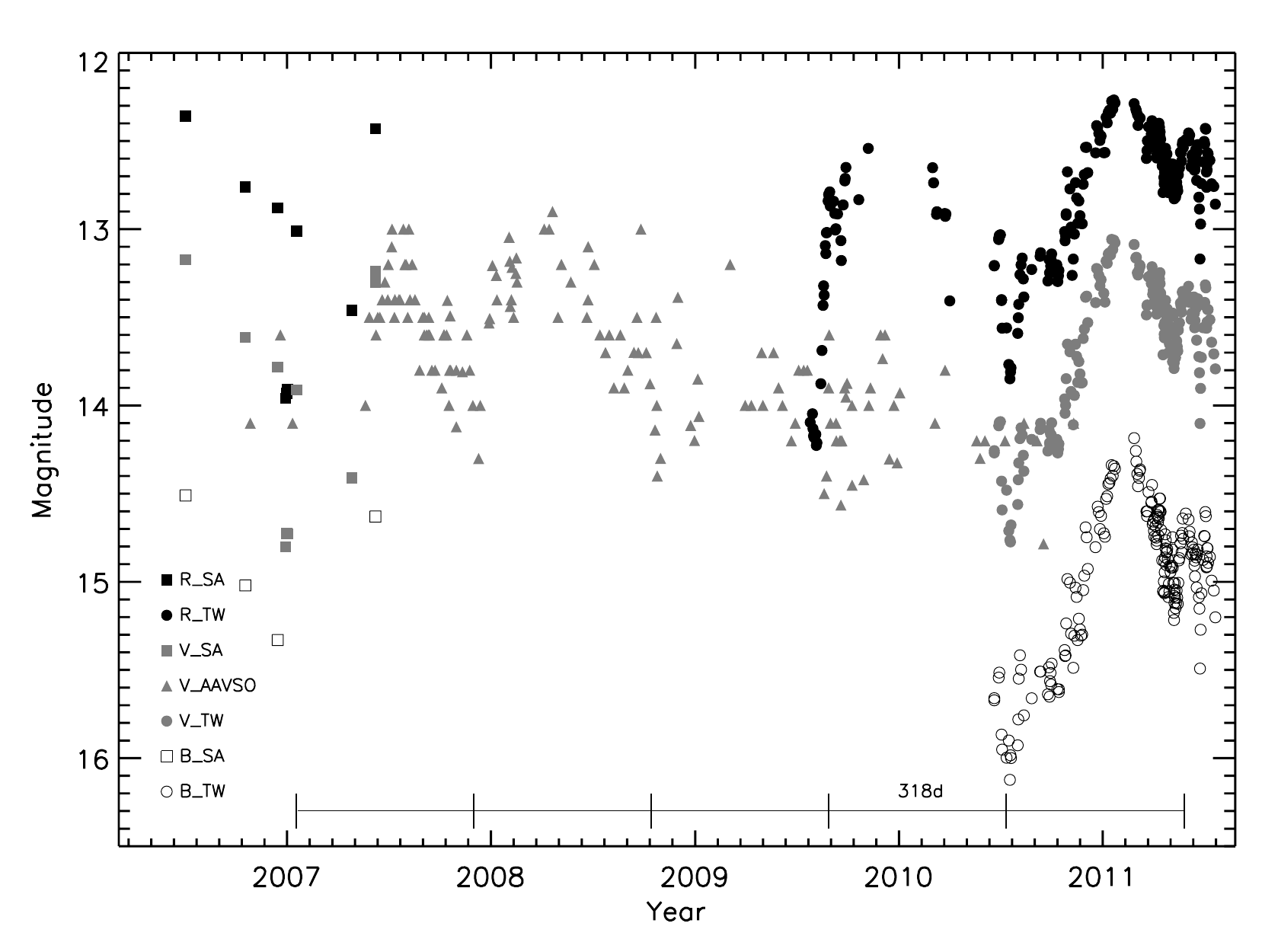}
\caption{The long term light curves of GM\,Cep in the $B$ (open symbols), $V$ (gray symbols), and $R$ (filled symbols) bands.  The data are from \citet{sic08} (squares), AAVSO (triangles), and our own observations (circles). \label{fig3}}
\label{fig3}
\end{center}
\end{figure}

\section{Discussions}
The sporadic brightening episodes accompanied with a blue color can be accounted for by an enhanced accretion activity.
The fading---a duration of 39~d with a blue color, and a period of 311~d---is puzzling.
A possible mechanism could be due to obscuration by a clump of dust around the star, i.e., as suggested for
UXor-type variables \citep{her94}.  We therefore suggest GM\,Cep to be an UXor, rather than
an EXor \citep{sic08}.

Assuming the clump is in Keplerian motion, and given the stellar mass of $2.1~M\sun$ and an orbital period of $P=311$~d, we derive the orbital distance of
$a\sim 1.16$~AU.  From the duration of the fading episode $t$, we have
\begin{equation}
\frac{t}{P}=\frac{2R_C}{2 \pi a},
\end{equation}
\noindent
where $t=38.9$~days, and $R_C\sim0.47$~AU is the radius of the clump.  The clump has a size of about $100~R\sun$.

Furthermore, the amount of fading allows us to estimate the amount of obscuring dust along the line of sight, i.e.,
$A_{\lambda}  = 1.086 \, N_d \, \sigma_d \, Q_{ext}$, where $N_d$ is the column density of the dust grains,
$\sigma_d$ is the geometric cross section of a grain, and $Q_{ext}$ is the dimensionless extinction efficiency factor.   
For particle sizes comparable to the wavelength, $Q_{ext}\sim 1$.
Stars of a few Myr old would have the large grains
settled into the midplane, so we assume the obscuration is attributed mostly to dust grains with an average radius of
$\sim0.1$~\micron.   Given the observed $A_V  = 0.68$~mag, the column density of dust then is
$ N_d = 1.994 \times 10^7 \,\mbox {cm}^{-2}$, which leads to a volume density of
$ n_d = N_d / 2R_C = 1.424 \times 10^{-6}\, \mbox {cm}^{-3}$.

Since the clump is quite close to the star ($a=1.16$~AU), we assume the composition of the dust clump is mostly silicates, having an
average density of $ \rho=3.5~$g\,cm$^{-3}$.  Therefore we derive the mass of the clump to be

\begin{equation}
M_d={\frac{4}{3}}\pi  R^3 n_d m_d = 2.269 \times 10^{21}\,\mbox{g} = 2.269 \times 10^{18}\,\mbox{kg,}
\end{equation}

\noindent
which is about that of an asteroid.

About a dozen UXors have been known thus far.  Some show cyclic variability, with periods ranging from 8.2~days
\citep{bou03} to 11.2~years \citep{gri98}.  We propose that GM\,Cep belongs to this category of objects experiencing
highly variable circumstellar extinction.  The mass we derived for the clump is only for the dust, and we have no
evidence, even with a sufficient amount of associated gas, if the clump is on the verge of gravitational instability.
Apparently a relatively small amount of dust mass along the line of sight to the circumstellar disk could cause the
obscuration event we detected in our light curve.  The intervening material does not need to shape like a clump, as long as
the material is not evenly distributed azimuthally, e.g., a warped or arm-spiraled disk.  Such inhomogeneity in the young
stellar disk might signpost
the continuing process from grain growth to the onset of planetesimal formation.
Further characterization of the ununiform disk of GM\,Cep, e.g., by polarization, reoccurrence of the bluing phenomenon in the fading epoch,
infrared spectroscopy and submillimeter imaging in and out of the obscuration phase would shed more light on our hypothesis.

\acknowledgements  The work reported here is particularly supported by the National Science Council grant NSC99-2119-M-008-021.  Hu thanks
the hospitality of the Yunnan Observatory LOC during the conference.

\bibliography{author}

\end{document}